# Spatial Analysis of Seasonal Precipitation over Iran: Co-Variation with Climate Indices

**Majid Dehghani [1], Somayeh Salehi [2], Amir Mosavi [3,4], Narjes Nabipour [5,*], Shahaboddin Shamshirband [6,7,*], and Pedram Ghamisi [8]**

1 Department of Civil Engineering, Technical and Engineering Faculty, Vali-e-Asr University of Rafsanjan, P.O. Box 518, Rafsanjan, Iran; m.dehghani@vru.ac.ir
2 Department of Civil Engineering, Technical and Engineering Faculty, Shahid Bahonar University of Kerman, Kerman, Iran; somayehsalehi @eng.uk.ac.ir
3 School of the Built Environment, Oxford Brookes University, Oxford OX3 0BP, UK; a.mosavi@brookes.ac.uk
4 Kalman Kando Faculty of Electrical Engineering, Obuda University, Budapest, Hungary; amir.mosavi@kvk.uni-obuda.hu
5 Institute of Research and Development, Duy Tan University, Da Nang 550000, Vietnam; narjesnabipour@duytan.edu.vn
6 Department for Management of Science and Technology Development, Ton Duc Thang University, Ho Chi Minh City, Vietnam; shahaboddin.shamshirband@tdtu.edu.vn
7 Faculty of Information Technology, Ton Duc Thang University, Ho Chi Minh City, Vietnam
8 Exploration Division, Helmholtz Institute Freiberg for Resource Technology, Helmholtz-Zentrum Dresden-Rossendorf, Dresden, Germany;

**Abstract:** Temporary changes in precipitation may lead to sustained and severe drought or massive floods in different parts of the world. Knowing variation in precipitation can effectively help the water resources decision-makers in water resources management. Large-scale circulation drivers have a considerable impact on precipitation in different parts of the world. In this research, the impact of El Niño-Southern Oscillation (ENSO), Pacific Decadal Oscillation (PDO), and North Atlantic Oscillation (NAO) on seasonal precipitation over Iran was investigated. For this purpose, 103 synoptic stations with at least 30 years of data were utilized. The Spearman correlation coefficient between the indices in the previous 12 months with seasonal precipitation was calculated, and the meaningful correlations were extracted. Then the month in which each of these indices has the highest correlation with seasonal precipitation was determined. Finally, the overall amount of increase or decrease in seasonal precipitation due to each of these indices was calculated. Results indicate the Southern Oscillation Index (SOI), NAO, and PDO have the most impact on seasonal precipitation, respectively. Also, these indices have the highest impact on the precipitation in winter, autumn, spring, and summer, respectively. SOI has a diverse impact on winter precipitation compared to the PDO and NAO, while in the other seasons, each index has its special impact on seasonal precipitation. Generally, all indices in different phases may decrease the seasonal precipitation up to 100%. However, the seasonal precipitation may increase more than 100% in different seasons due to the impact of these indices. The results of this study can be used effectively in water resources management and especially in dam operation.

## 1. Introduction

The role of climate in human life and other organisms is crucial. Any seasonal and periodic fluctuations in the climate of an area can seriously affect living conditions. These fluctuations in arid and semi-arid regions of the world have more significant impacts[1-3]. Iran is located in an arid and semi-arid region with less than a third of the world average precipitation, which shows limited water resources in this country[4,5]. The precipitation varies significantly between the years, which makes a challenge for water resources management. A countrywide drought that occurred in 2018 in Iran led to about a53% decrease in precipitation compared with the 49-year average[6]. On the contrary, 50% increase in precipitation compared with the 50-year average occurred in 2019 resulted in considerable floods in Northern and southwestern parts of Iran (Iran's ministry of energy reports). On the other hand, the population of the country in 1986 (30 years ago) was approximately 50 million people which grows to almost 80 million in 2016. Increasing the population, limited water resources, and precipitation variability necessitate complicated water resources management. One of the earliest needs for efficient water resources planning and management is the prior knowledge about precipitation. While long-term exact precipitation forecasting is difficult, information about the increase or decrease in precipitation can help the water resources policymakers. The atmospheric-ocean interactions can help to find mechanisms that affect the distribution and magnitude of precipitation in a region[5]. Several researchers have found a relationship between large-scale circulation drivers and temporary climate variations[7-20]. Among all, the El Niño-Southern Oscillation (ENSO), Pacific Decadal Oscillation (PDO), and North Atlantic Oscillation (NAO) are distinguished as the most essential phenomena due to its considerable impacts on hydro-climatic systems.

Several studies were conducted to find the impact of these signals on precipitation in Iran. Nazemosadat and Cordery [21] evaluated the impact of SOI on the autumn rainfall in Iran. They reported a negative relationship between the SOI and autumn rainfall in almost all parts of Iran. Also, the impacts of SOI on rainfall amounts during El Niño and La Nina phases were studied, which indicates that the amount of precipitation highly intensifies during the El Niño. Nazemosadat and Ghasemi [22] assessed the impact of SOI on dry and wet periods in Iran. It has been found that El Niño increases the autumn precipitation in the southern parts of the country while La Nina reduces the amount of precipitation. Raziei et al.[23] analyzed the drought in western Iran from 1966 to 2000 by the Standardized Precipitation Index (SPI) and reported that they could not find a substantial correlation between drought and wet periods with the SOI. Tabari et al.[24]reported that the relationship between the North Atlantic Oscillation (NAO) and the seasonal and annual streamflow in the west of Iran was statistically weak and insignificant during the last four decades. Biabanaki et al.[5] evaluated the variability of precipitation in western Iran using the SPI. They reported SOI and PDO strongly affect the precipitation in western Iran. Roghani et al.[25] investigated the impact of SOI on autumn rainfall in Iran and showed the persistence of Southern signals on autumn rainfall. Alizadeh-Choobari et al.[26] investigated climate variability from 1980 to 2016 using EP/CP ENSO signals. Results indicated that the ENSO cycle contributes to the inter-annual climate variability over Iran. Alizadeh-Choobari and Najafi [27] investigated the precipitation in Iran in response to the El Niño-Southern Oscillation. The complicated interaction between ENSO and seasonal precipitation, especially in winter and spring. Several studies [28-30] analyzed the covariation of climate indices PDO, NAO, and SOI with precipitation to assess to assess meteorological and hydrological drought in Iran. Also, in other parts of the world, some research were carried out on the relationship between climate indices and hydrological variables[31,32].

It is apparent that large-scale circulation drivers cause precipitation variability in Iran with different intensities. However, there are several undefined aspects that must be clarified. In previous studies, most attention has given to the relation of SOI with the precipitation, while other signals such as PDO and NAO were rarely considered for seasonal precipitation. Second, the amount of decrease or increase in precipitation compared to the long-term average was not reported in the previous research. On the other hand, most of the previous studies focused on autumn and winter rainfalls. Salehi et al. [4] reported an increase and decrease in spring and winter precipitation in Iran, respectively. So, it seems that it is needed to explore the relationship of large-scale circulation drivers with the precipitation in all seasons. These shortcomings in previous research motivated the authors to clarify these aspects. To do this, in the present study, the relationship of seasonal precipitation with SOI, NAO, and PDO was investigated. For this purpose, 103 synoptic stations with at least 30 years of data are used. Also the percentage of increase/decrease in precipitation compared with the long-term average is calculated. The rest of the paper is organized as follows. The study area, SOI, PDO, NAO, and the methodology of research are presented in the next section. The results of the study and associated discussion are explained in details in the results section. Finally, the conclusions are drawn in the last section.

## 2. Materials and Methods

### *2.1. Study area*

Iran is a large country in southwest Asia between about 25° $N$ to 40° $N$, and 45° $E$ to 62° $E$with a total area of 1648195 km2. It is located between the Caspian Sea in the north and the Persian Gulf and Oman Sea in the south. Iran is located in an arid and semi-arid region with less than one-third of precipitation compared to the world's annual average. There are two great mountain chains, including Alborz and Zagros in Iran. Alborz mountain chain stretches from west to east below the Caspian Sea while Zagros stretches from northwest to southeast of Iran (Figure 1). The main basins (six basins) and the main catchments (30 catchments) are plotted in Figure 1.

### *2.2. Data*

#### 2.2.1. Rainfall

In this study, 103 synoptic stations are selected, which have at least 30 years of qualified monthly precipitation data which covers the whole country. These stations were chosen among 220 stations. Trend, homogeneity, and randomness of the data were examined for all 220 stations and these 103 stations were selected based on these tests. The data covers the period from 1987 to 2016. The data were collected from the Islamic Republic of Iran Meteorological Office. The annual average of precipitation in these stations is presented in Table A1 in the Appendix. The annual average of precipitation varies from 51mm in Yazd to 1730 mm in Bandar-e-Anzali.

#### 2.2.2.Climate indices: El Niño-southern oscillation

The Southern Oscillation Index (SOI) is defined as a standardized index based on the difference in sea level air pressure between Tahiti in the central South Pacific Ocean and Darwin in Australia. SOI shows the large-scale fluctuation in air pressure between east and west of tropical Pacific Nazemosadat and Cordery [21]. Negative values of the SOI show El Niño while positive values correspond to the La Niña phase. During the El Niño, the air pressure at Darwin is above normal while it is below normal in Tahiti. Based on the database used in this study, the SOI ranges from -4 to 4. Five SOI phases are considered based on the classification by Saghafian et al. [33] (Table 1).

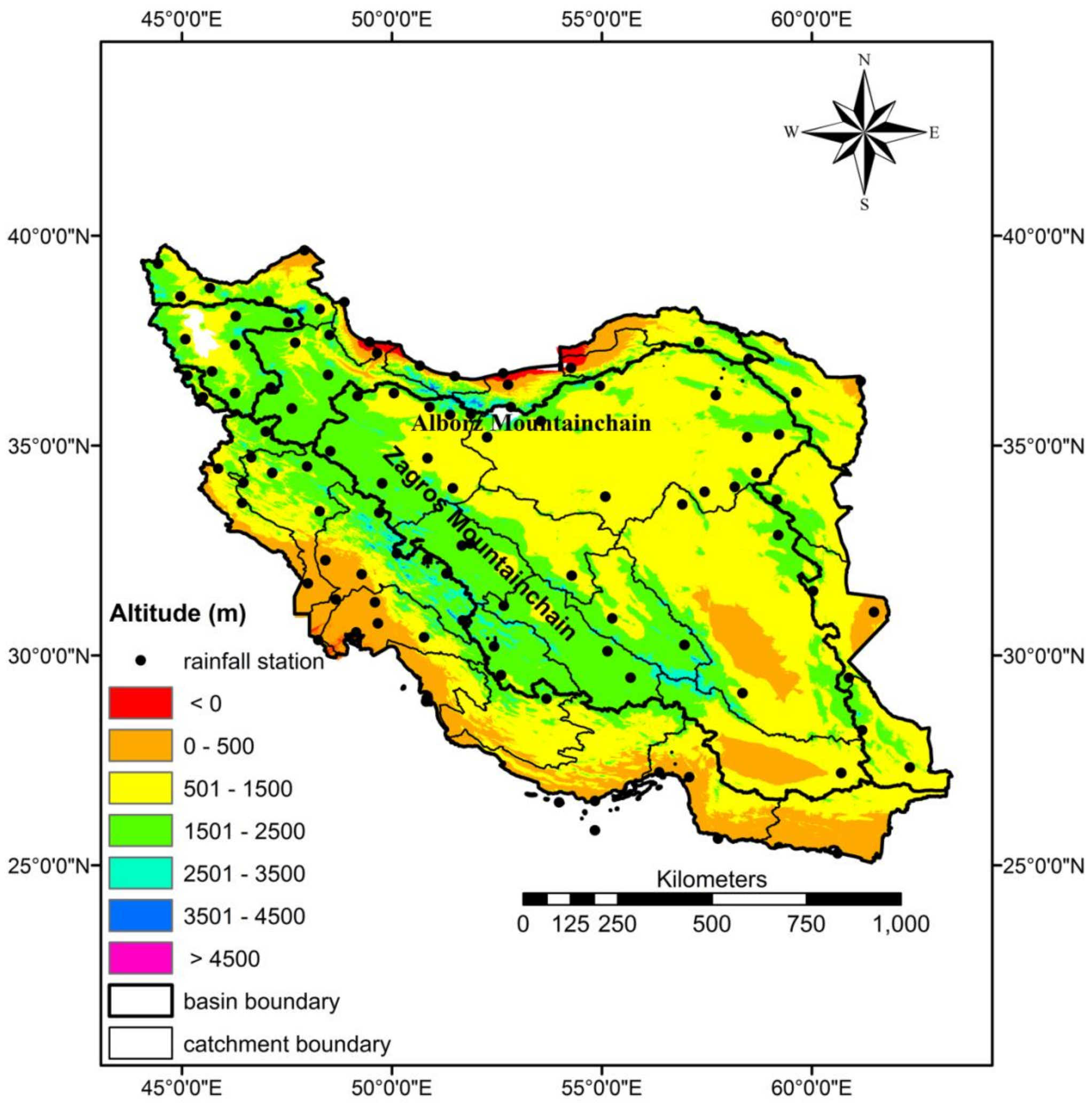

**Figure 1.** Location map of study region and synoptic stations.

**Table 1.** Large-scale climate circulation indices classification. The x denotes the value of each index.

| Range of indices | SOI | PDO | NAO |
|---|---|---|---|
| $x \leq -2$ | strong El Niño | Very warm | Very cold |
| $-2 \leq x < -1$ | El Niño | Warm | Cold |
| $-1 \leq x < 1$ | Neutral | Neutral | Neutral |
| $1 \leq x < 2$ | La Niña | Cold | Warm |
| $x \geq 2$ | strong La Niña | Very cold | Very warm |

2.2.3. Climate indices: North Atlantic oscillation (NAO)

The North Atlantic Oscillation (NAO) index is based on the difference between sea surface pressure in Azores and Iceland. As the impact of the NAO is regional, the NAO is defined as a normalized index which is calculated based on the difference between sea-level pressure at Azores and Iceland (Cullen et al., [34]). This fluctuation has a strong impact on the atmospheric pattern, especially in winter in different parts of the world, including in North America, North Africa, northern Asia, and Europe (Tabari et al., [24]). In the positive or cold phase, intense high pressure is centered over the Azores while an intense low pressure is located over Iceland. The negative phase of NAO happens when both sub-polar low pressure and subtropical high pressure are below the average. In this research, the classification of NAO was proposed the same as SOI.

2.2.4. Climate indices:Pacific Decadal Oscillation (PDO)

Pacific decadal oscillation (PDO) is a sea surface temperature (SST) anomaly in the Pacific Ocean. In the negative or cold phase, SST is cool and extends from the equator into the Gulf of Alaska [35]. In contrast, during the positive or warm phase, the warm SST extends from the equator into the Gulf of Alaska. The PDO phases are similar to SOI. Also, the classification of PDO was proposed the same as SOI. The SOI, PDO, and NAO data were collected from the National Oceanographic and Atmospheric Administration (NOAA) (data available at https://www.esrl.noaa.gov/psd/gcos_wgsp/Timeseries).

*2.3. Spatial characteristics of rainfall for the various seasons*

The long-term average of precipitation and coefficient of variation in different seasons were presented in Figure 2. According to this figure, the maximum precipitation occurs in winter and in the next step in autumn. Also, the western and especially the coastal regions in the north receive much more precipitation in all seasons compared to the other parts of the country. There is generally a diagonal reduction in precipitation from northwest to southeast in almost all seasons. Based on Figure 2, the coefficient variation of precipitation during winter is less than 100%. The country can be divided into northern and southern parts with less and more than 50% of the coefficient of variation, respectively. in the autumn and spring, the coefficient of variation in the south and southeast parts is more than 100% while in the other regions it is less than 100%. The highest coefficient of variation could be found in summer precipitation where the coefficient of variation in almost all parts of the country is more than 100%.

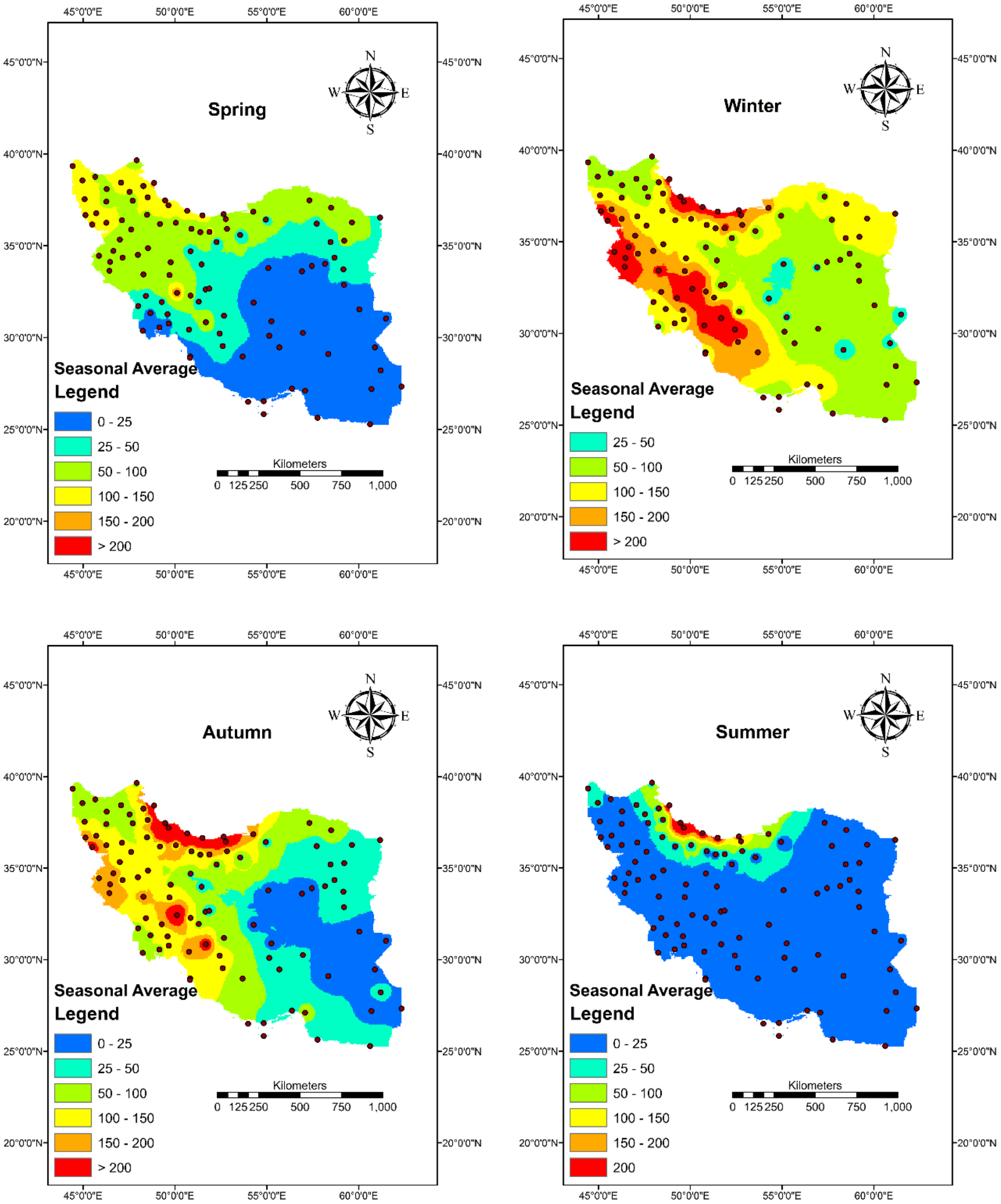

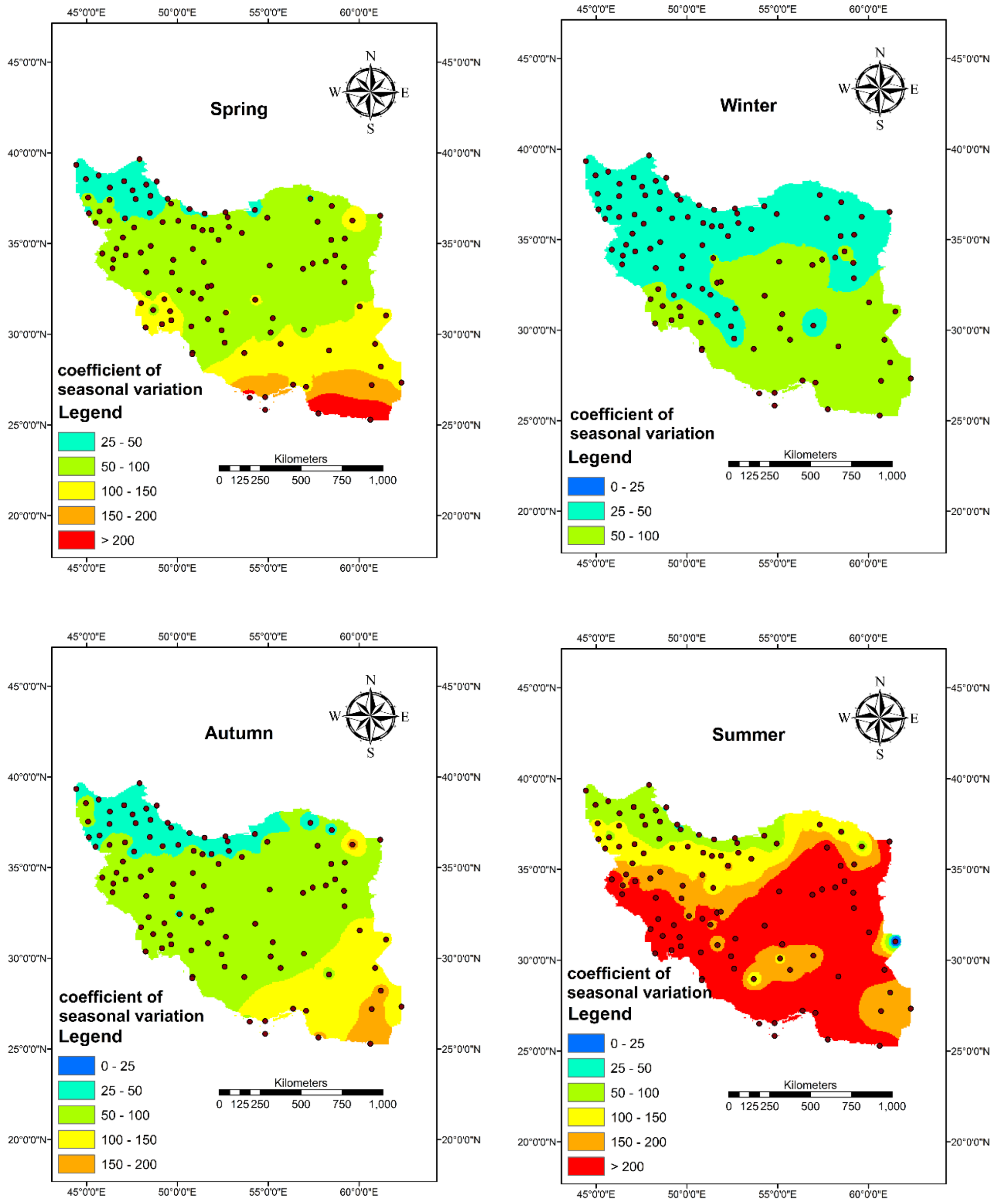

**Figure 2.** The long-term average of the precipitation and coefficient of variation (in percent) for Winter, Spring, Summer, and Autumn

## *2.4. Analyses of the correlation between rainfall and climate indices*

In this research, the impact of large-scale climate circulation indices on seasonal precipitation in Iran was investigated. For this purpose, 103 synoptic stations that had at least 30 years of data were selected, and the seasonal precipitation was calculated based on the monthly precipitation records. Then, the authors considered that the seasonal precipitation is correlated with monthly SOI, PDO and NAO in the previous 12 months. So,12 Pearson correlation coefficient was calculated between the monthly SOI, NAO, and PDO indices in the previous 12 months with the seasonal precipitation according to equation (1):

$$\rho_{X,Y} = \frac{cov(X,Y)}{\sigma_X \, \sigma_Y} \tag{1}$$

where $X$ and $Y$ denote the precipitation and the climate indices, respectively. Furthermore, $\rho$ is the Pearson correlation coefficient and σ is the standard deviation. The covariance can be calculated as:

$$cov(X,Y) = E[(X - \mu_X)(Y - \mu_Y)] \quad (2)$$

where $E$ and $\mu$ are the expectation and mean, respectively. Also, the $p$-value at $\alpha = 0.05$ significance level was calculated. The null hypothesis is that , where there is no correlation or dependence between the climate indices and seasonal precipitation. In contrast, the alternative hypothesis is that , and the correlation is meaningful. So, the $p$-value of less than 0.05 shows a meaningful correlation. In the next step, among the 12 calculated correlations, the one with the highest value (positive or negative) which has a $p$-value less than 0.05 is selected. The highest correlation shows that which month among the 12 months has the greatest effect on the seasonal precipitation. Then, the time series of SOI, NAO, and PDO in that month were categorized as presented in Table 1. In each category, the seasonal precipitation compared to the long-term seasonal precipitation and the total amount of decrease or increase in precipitation was calculated

**3. Results**

In this section, the results of the research were presented. First, the correlation coefficient between the climatic indices and seasonal precipitation was calculated and the results presented in Figures 3 to 5. Figure 3 shows the results of statistical test for correlation between seasonal precipitation and climatic indices, Based on the results, at $\alpha = 0.05$, SOI, PDO, and NAO have a significant relationship with winter precipitation in 72, 53, and 51 stations, respectively, while they have a significant relationship with spring precipitation in 74, 16, and 51 stations, respectively. It shows that all indices have a significant impact on winter precipitation in at least half of the stations. However, PDO has a negligible impact on spring precipitation. The weakest impact of these indices is in the summer. Just 30, 25, and 47 stations had a meaningful correlation with SOI, PDO, and NAO, respectively. Finally, in the autumn, 89, 33, and 38 stations had a meaningful correlation with SOI, PDO, and NAO, respectively. It can be concluded that the NAO affects the precipitation in about half of the stations during the winter, spring, and summer but has less impact on autumn precipitation. PDO has the most substantial impact on winter precipitation, and in other seasons, it has a negligible impact. However, SOI is the most important index with a strong influence on winter, spring, and especially the autumn precipitation. However, at $\alpha = 0.01$, NAO has significant relationship with a few stations especially in autumn and spring in the west regions of Iran. Also, PDO has significant relationship with just winter precipitation in the west regions in the country. In other seasons, there is no specific pattern and few stations with significant correlation scattered in the country. In contrast, SOI has significant correlation with spring and autumn precipitation in majority of stations. This significant correlation can be detected in the southern part of the country in winter while most of the stations in northern part of the country has no meaningful correlation even at $\alpha = 0.05$.

The general pattern of Spearman correlation between the NAO, PDO, SOI, and seasonal precipitation is drawn in Figure 4. For NAO, in winter, the north and southeast of the country have a negative correlation with NAO and the highest correlation indicated in the east of the country and some other stations in other parts. In spring, it is possible to divide the country into three parts. North, northwest, south and some parts in the middle of the country show a negative correlation with NAO while northeast and west of the country show a positive correlation. There is no meaningful correlation with NAO in the southwest. In the summer, the general pattern shows that almost all parts of the country have a negative correlation with NAO. In the autumn, north and central parts of the country have a negative correlation with NAO, while in the other regions, especially in the northwest, it is negative. There is no meaningful correlation in the southwest.

For PDO, in winter, the whole of the country shows a positive correlation with PDO except in one station in the west. In the spring, north, northwest, and northeast (except some stations in the northwest and northeast) show a positive correlation. South, southwest, and central parts of the country have no meaningful correlation with PDO, and the rest shows a negative correlation. In the summer, southeast, some regions in the southwest and west of the country represent a negative correlation, and the other regions follow a positive correlation with PDO in general. In autumn, southeast and northwest of the country have a positive correlation with PDO, and the rest of the country represents a negative correlation.

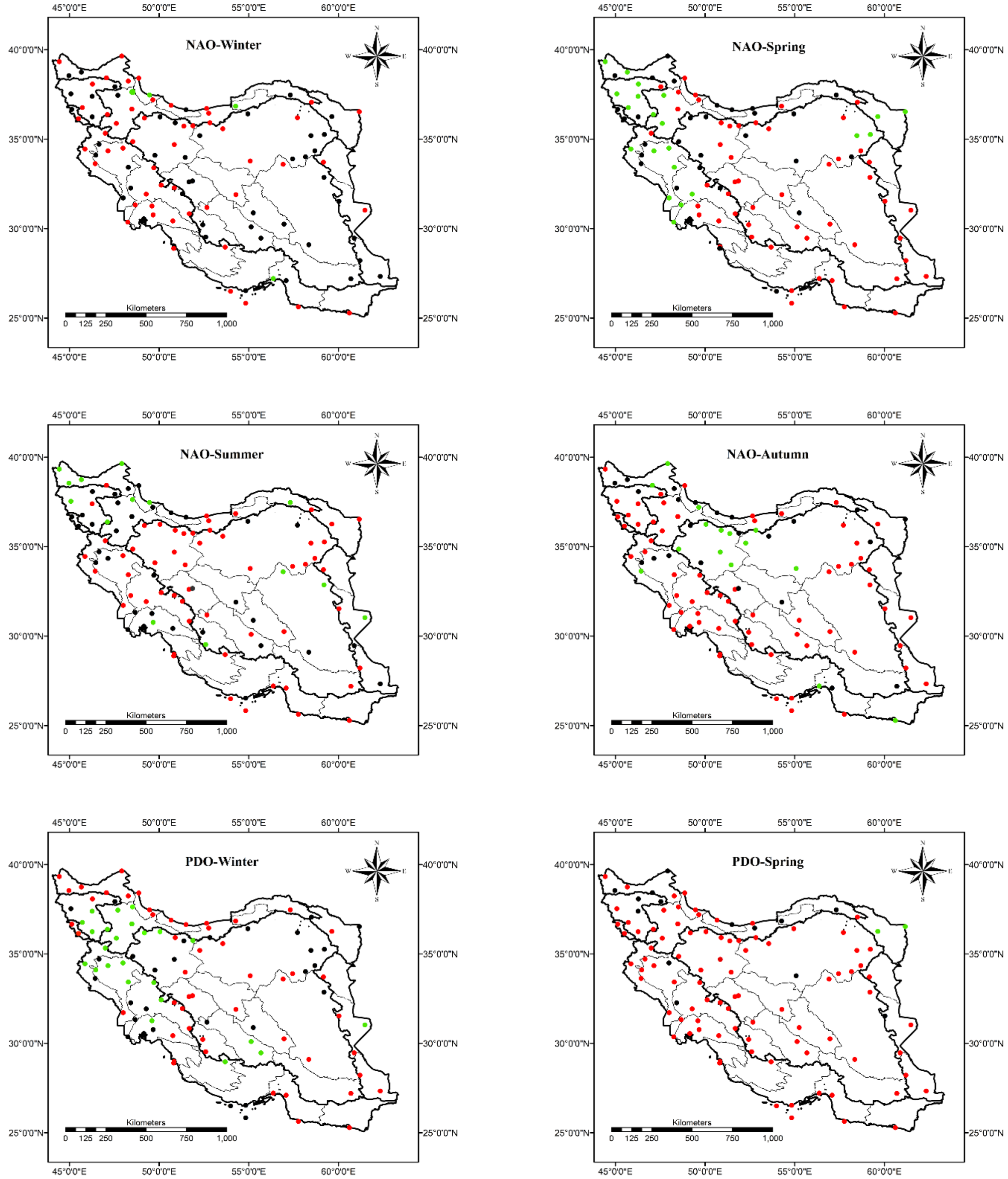

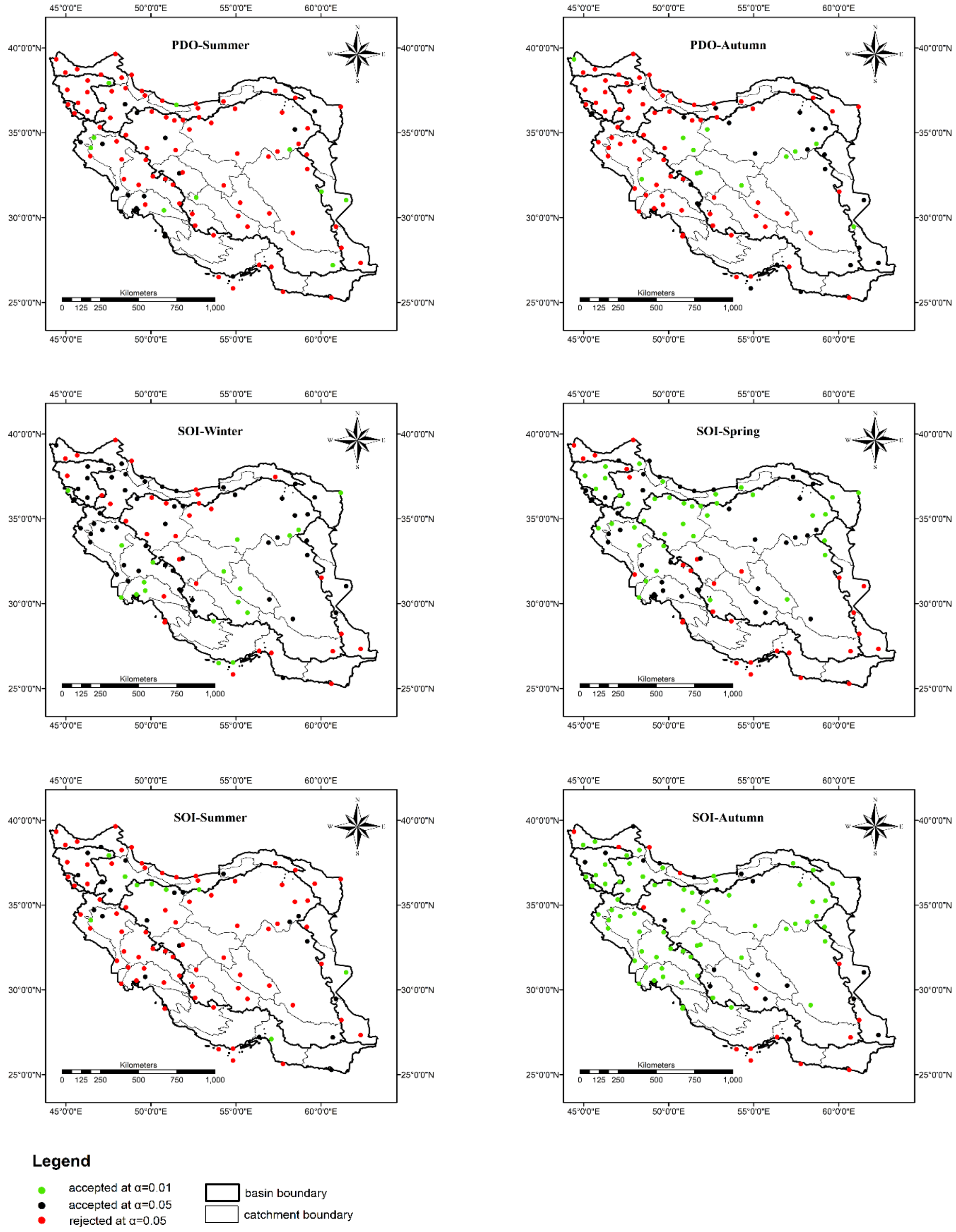

Figure 3. Results of statistical test for correlation between reasonal precipitation and climatic indices.

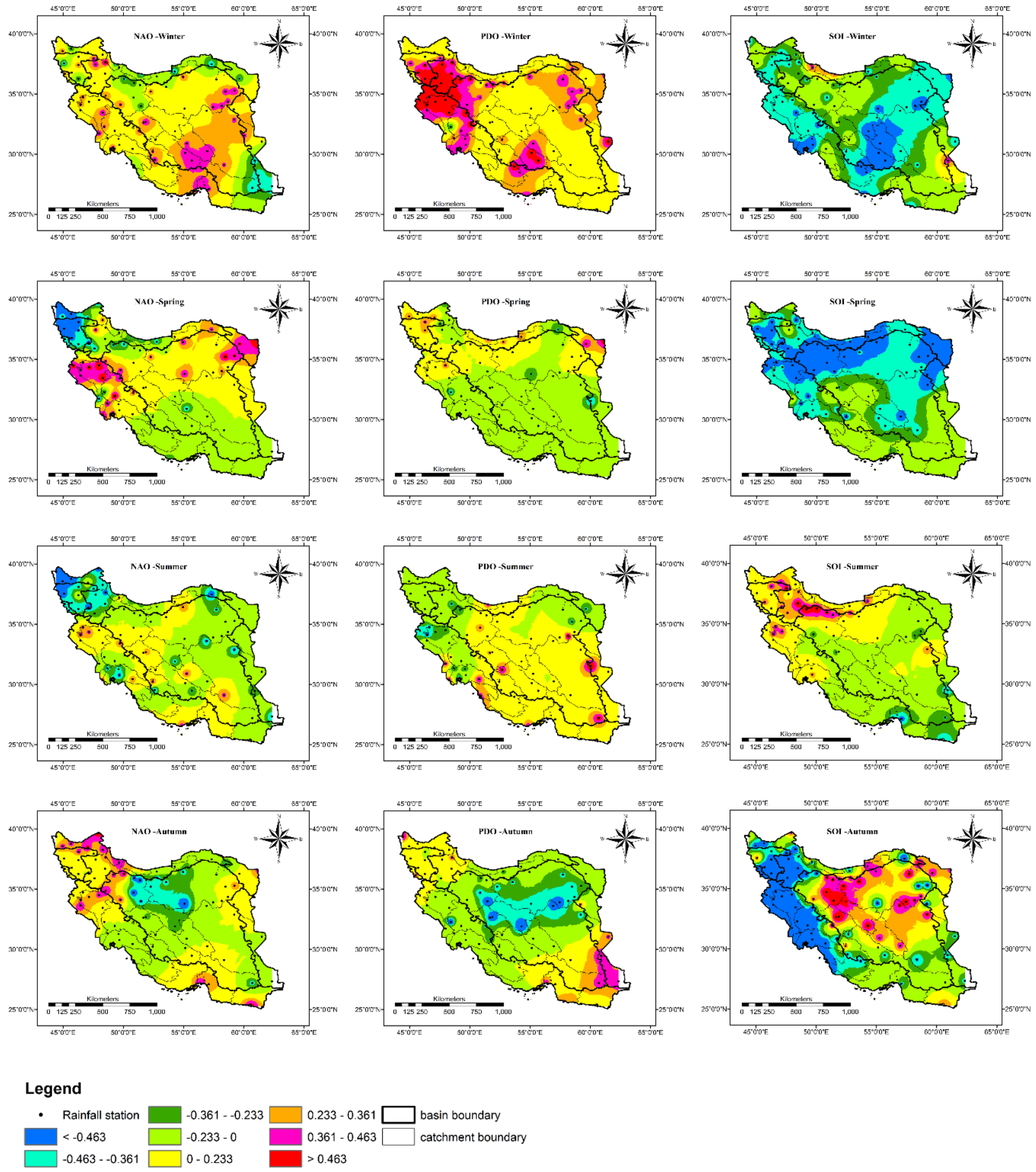

**Figure 4.** Spearman correlation between NAO, PDO, and SOI and seasonal precipitation.

For SOI, in winter, the whole country shows a negative correlation with SOI except in four stations, and this pattern continued to the spring. However, in the summer, north, west, and northwest of the country have a positive correlation with SOI, and the rest of the country shows a negative correlation. In the autumn, there is a high negative correlation with SOI in the west, northwest and it is dominant in south and southeast of the country with lower correlation. Other parts have a positive correlation with SOI. As the number of stations with meaningful correlation at $\alpha = 0.05$ is considerably higher than the case that $\alpha = 0.01$, so, the remaining analysis was done considering $\alpha = 0.05$. In Figure 5, the month in which the SOI, NAO, and PDO have the highest correlation with seasonal precipitation is shown. Based on this figure, for NAO in winter, the stations around the Zagros have the strongest relationship with NAO in July or August. In other regions, there is no general pattern. In the spring, the strongest relationship is with October and November in the west, northwest, and northeast while there is no relationship with NAO in east and southeast of the country.

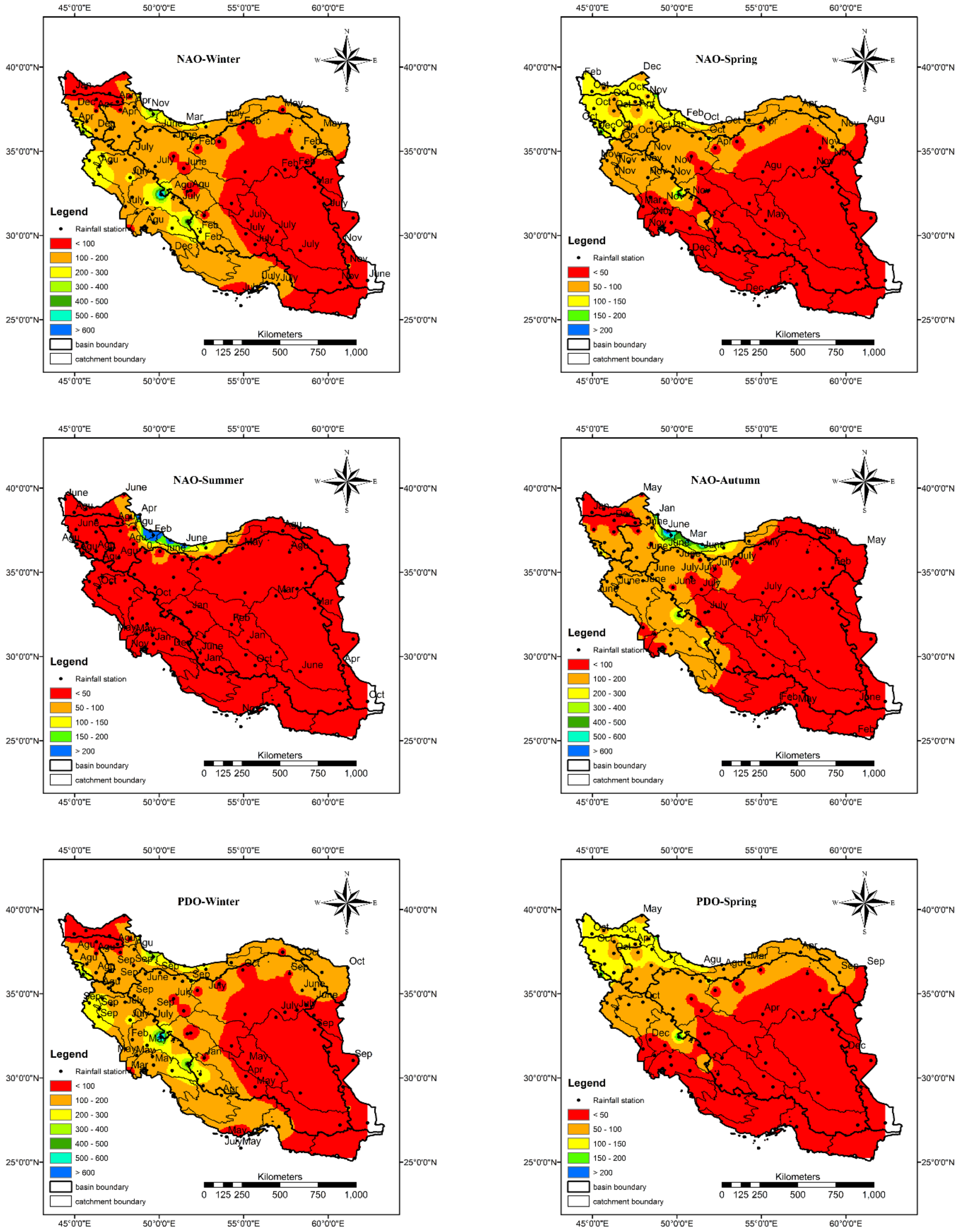

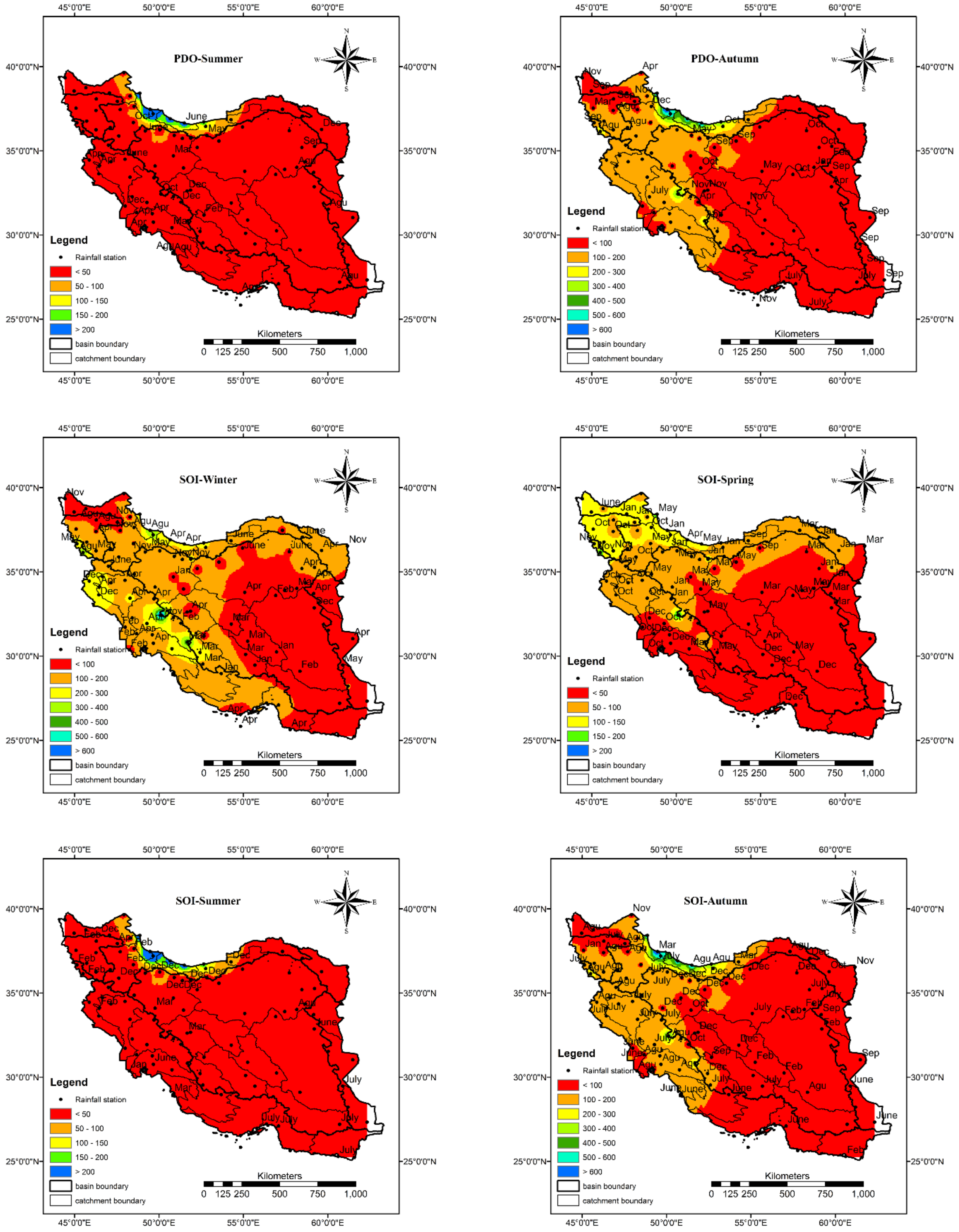

**Figure 5.** The month of SOI, PDO, and NAO with highest correlation with seasonal precipitation. The background is the long-term seasonal precipitation.

Finally, the average increase/decrease in precipitation due to the change in PDO, NAO, and SOI values is calculated and plotted in figures 6 to 10. Based on Figure 6, when the PDO, NAO, and SOI values are less than -2, different seasonal precipitation patterns can be detected. Generally in winter, when the PDO and NAO values are less than -2, the whole country encountered about 25% decrease in precipitation. Just north and especially southeast of the country may have above average precipitation associated with NAO<-2. However, SOI<-2 intensifies the precipitation in almost all parts of the country. It is considerable in the east and middle of the country. In the spring, all three indices intensify the precipitation in the southern half of the country, while PDO decreases the precipitation in northern parts and NAO in the middle and northeast. In general, SOI has the most significant impact on spring precipitation. In the summer, all indices increase the precipitation in the east of the country. However, SOI leads to a considerable decrease in precipitation in the northwest while NAO and PDO increase the precipitation in the northwest. Also, NAO and PDO lead to a significant decrease in precipitation in the middle of the country while SOI increases the precipitation. In the autumn, each index has a unique impact on precipitation. NAO decreases the precipitation in the northwest and southeast and an increase in other parts. PDO intensifies the precipitation in all parts of the country except in the southeast. SOI increases the precipitation in the northwest, southwest, and west, and some parts in the southeast and in the rest of the country, decreasing the precipitation is dominant.

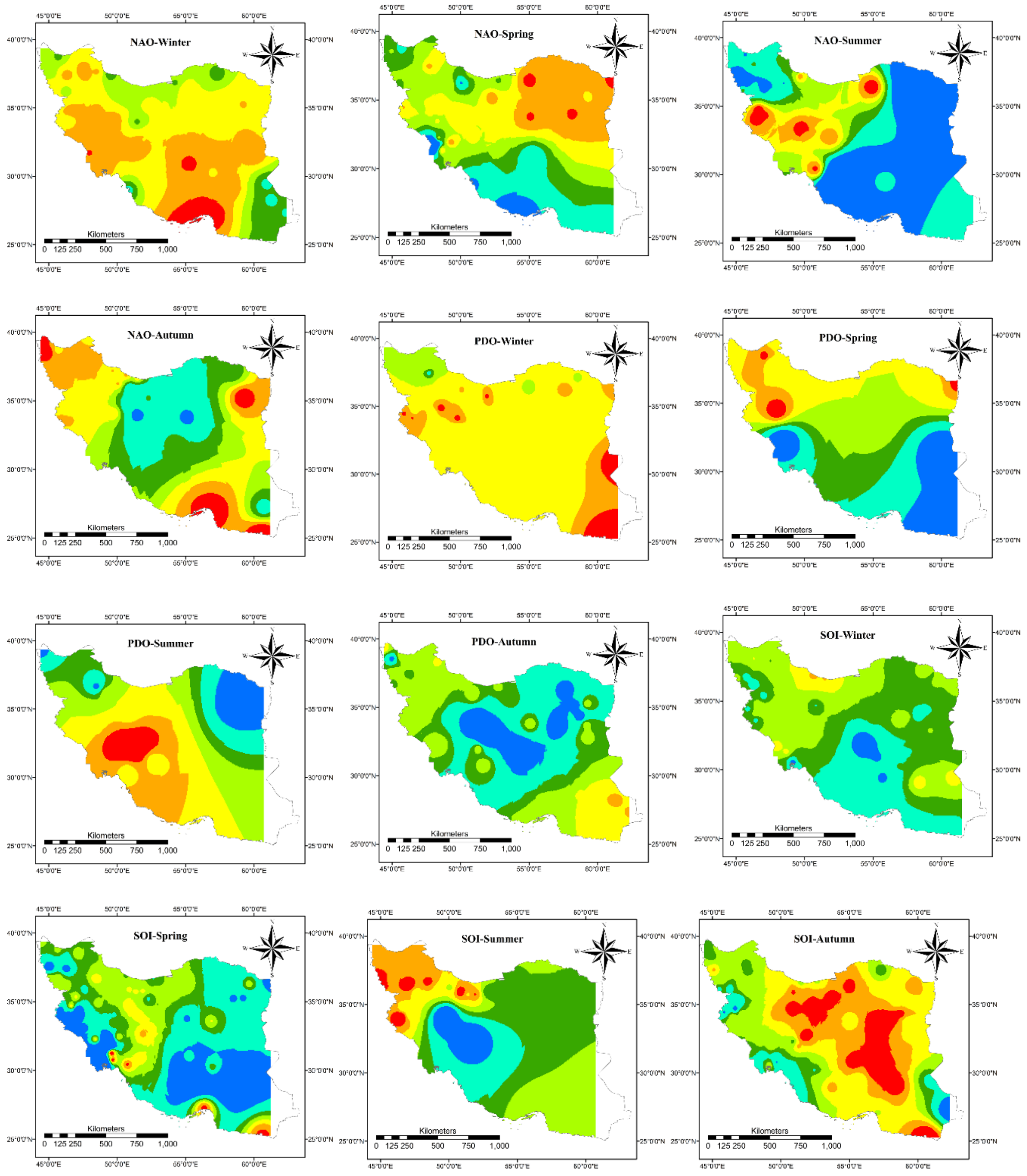

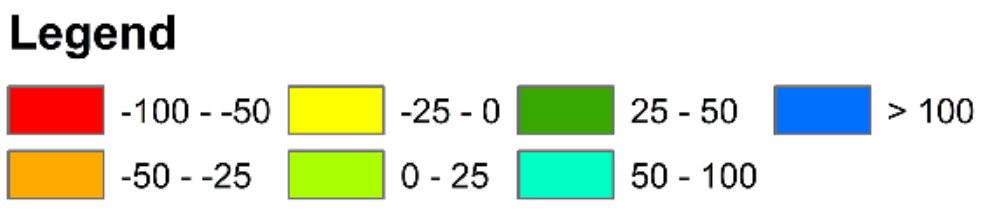

**Figure 6**. Total change in precipitation due to the effect of SOI, NAO, and PDO in the range of $x < -2$ where denotes the SOI, NAO, and PDO value.

Figure 7 shows the changes in precipitation when the NAO, PDO, and SOI are in the range of $-2 < x < -1$. The general pattern of precipitation in all seasons follows the results in Figure 6 with lower changes. However, there are some considerations. The overall change in winter precipitation associated with all indices not exceed from 50% except in some locations, especially in the southwest.

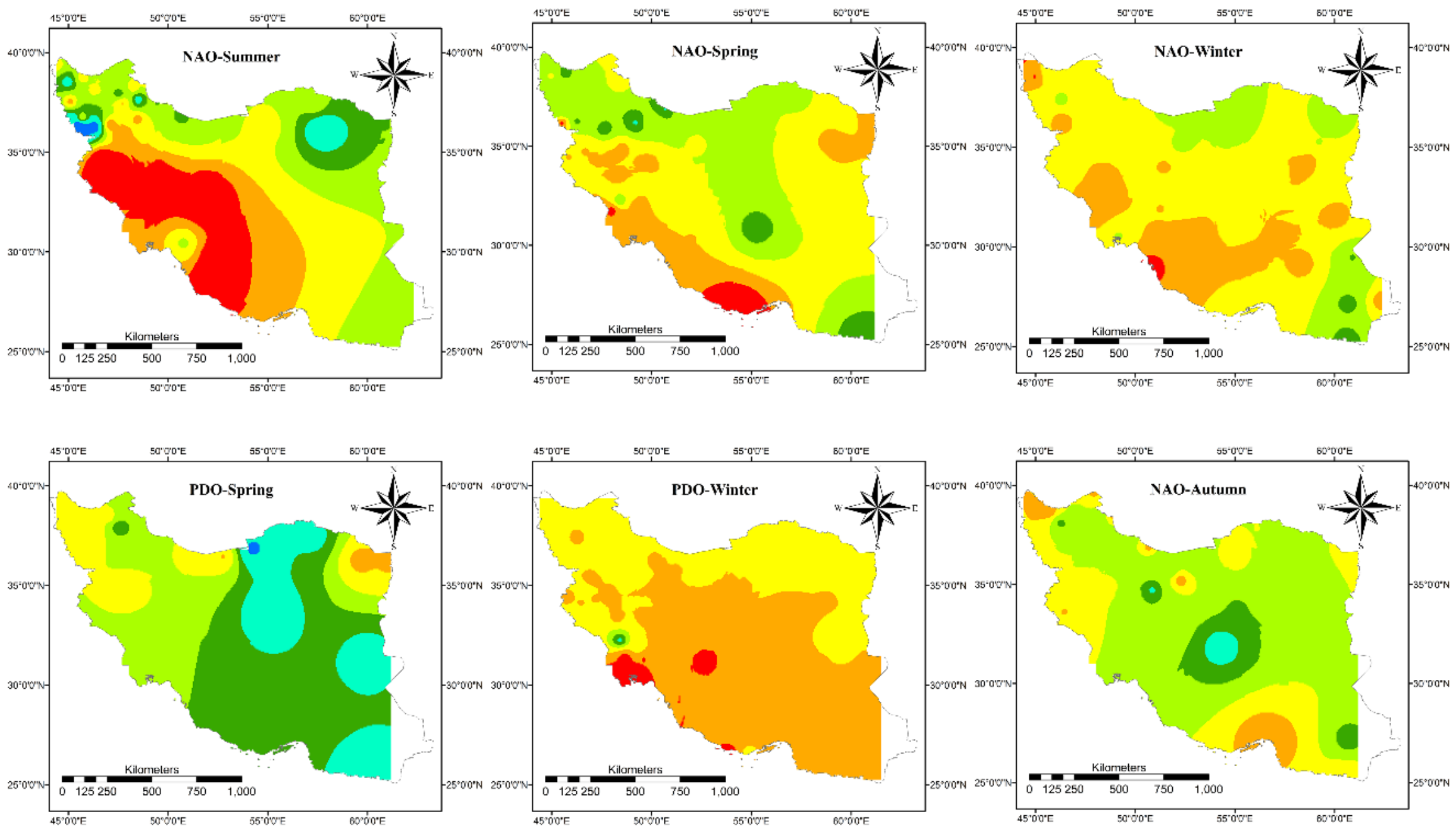

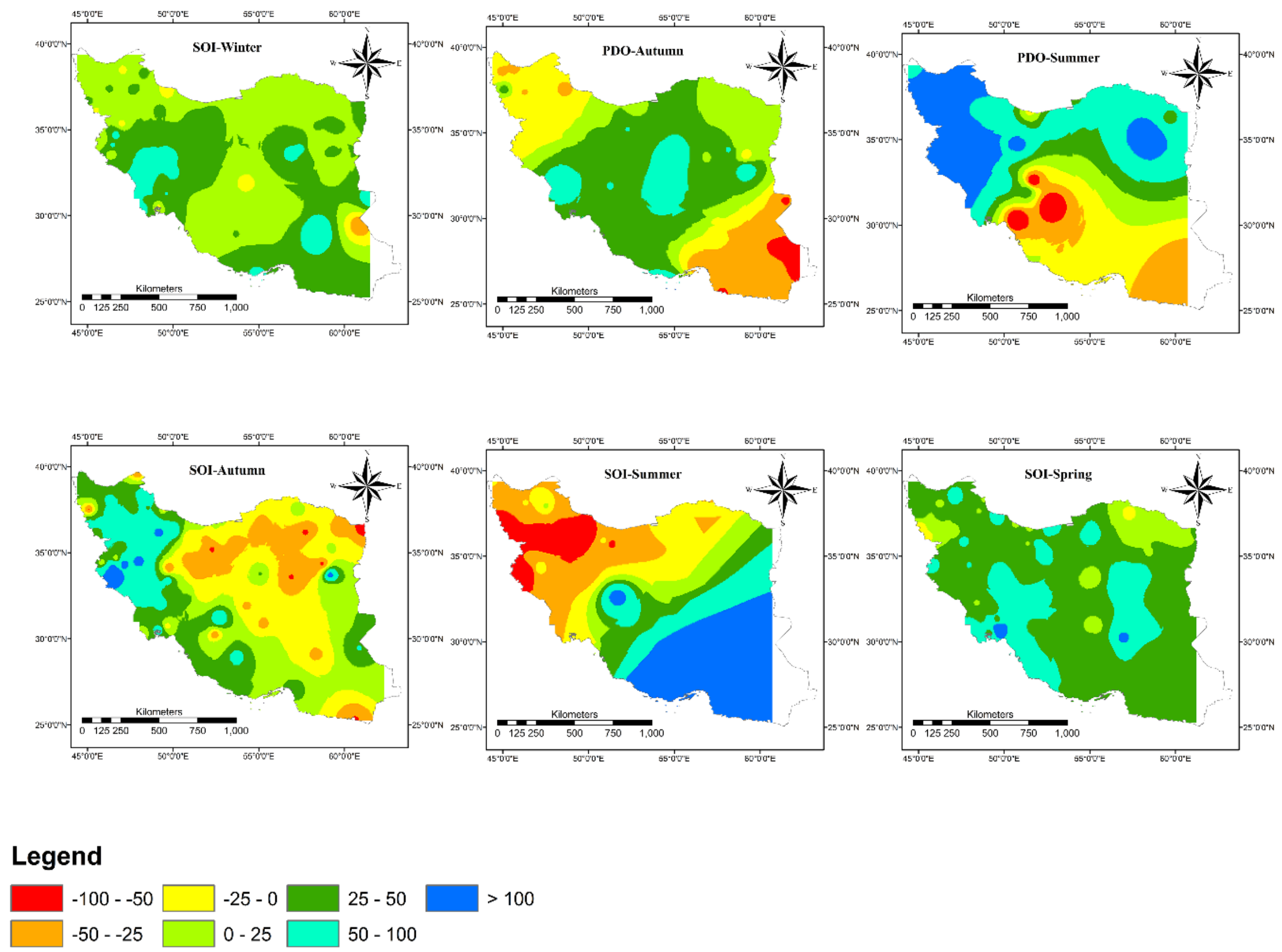

**Figure 7**. Total change in precipitation due to the effect of SOI, NAO, and PDO in the range of-2 <$x$ < -1 where $x$ denotes the SOI, NAO, and PDO value.

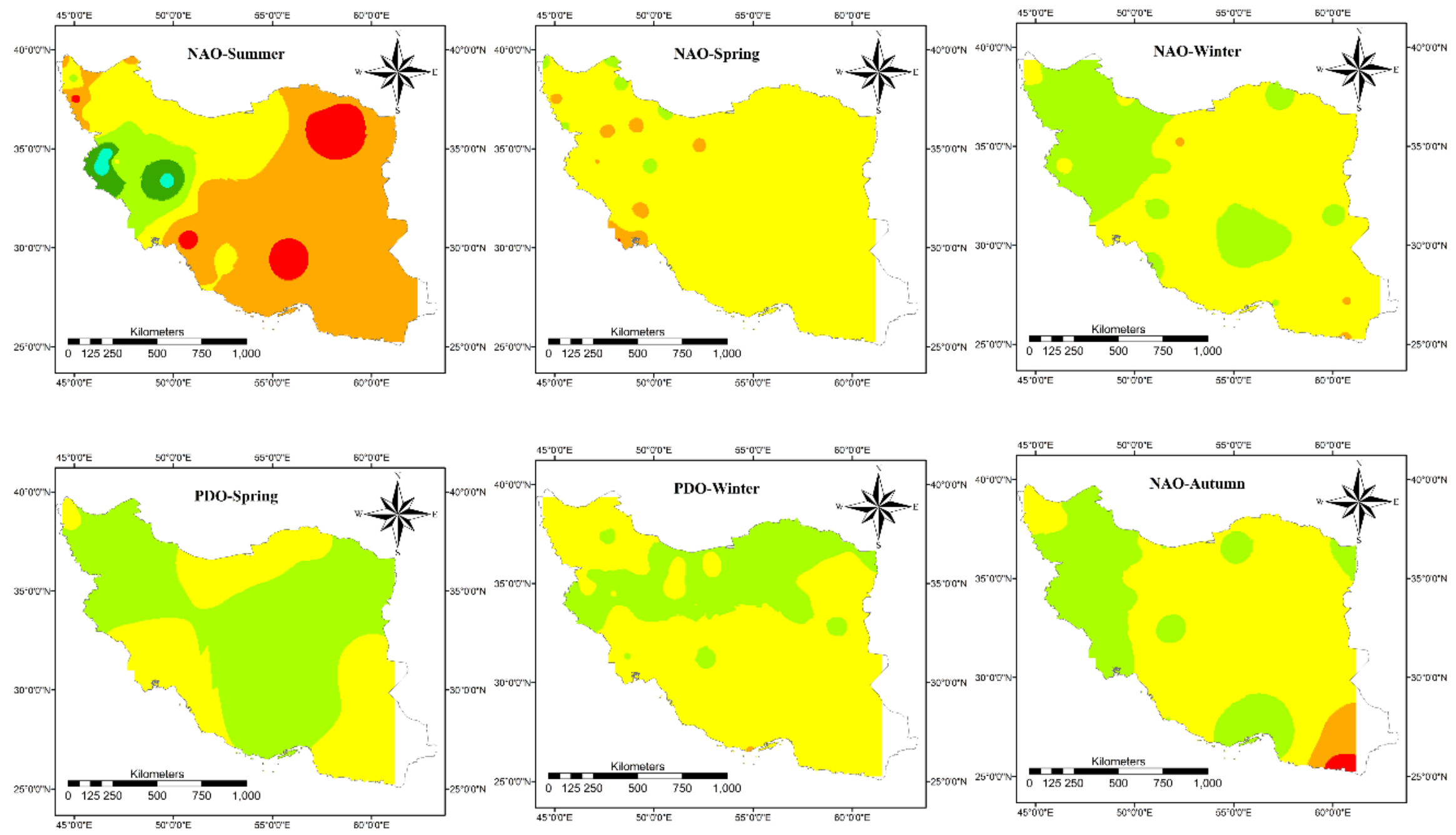

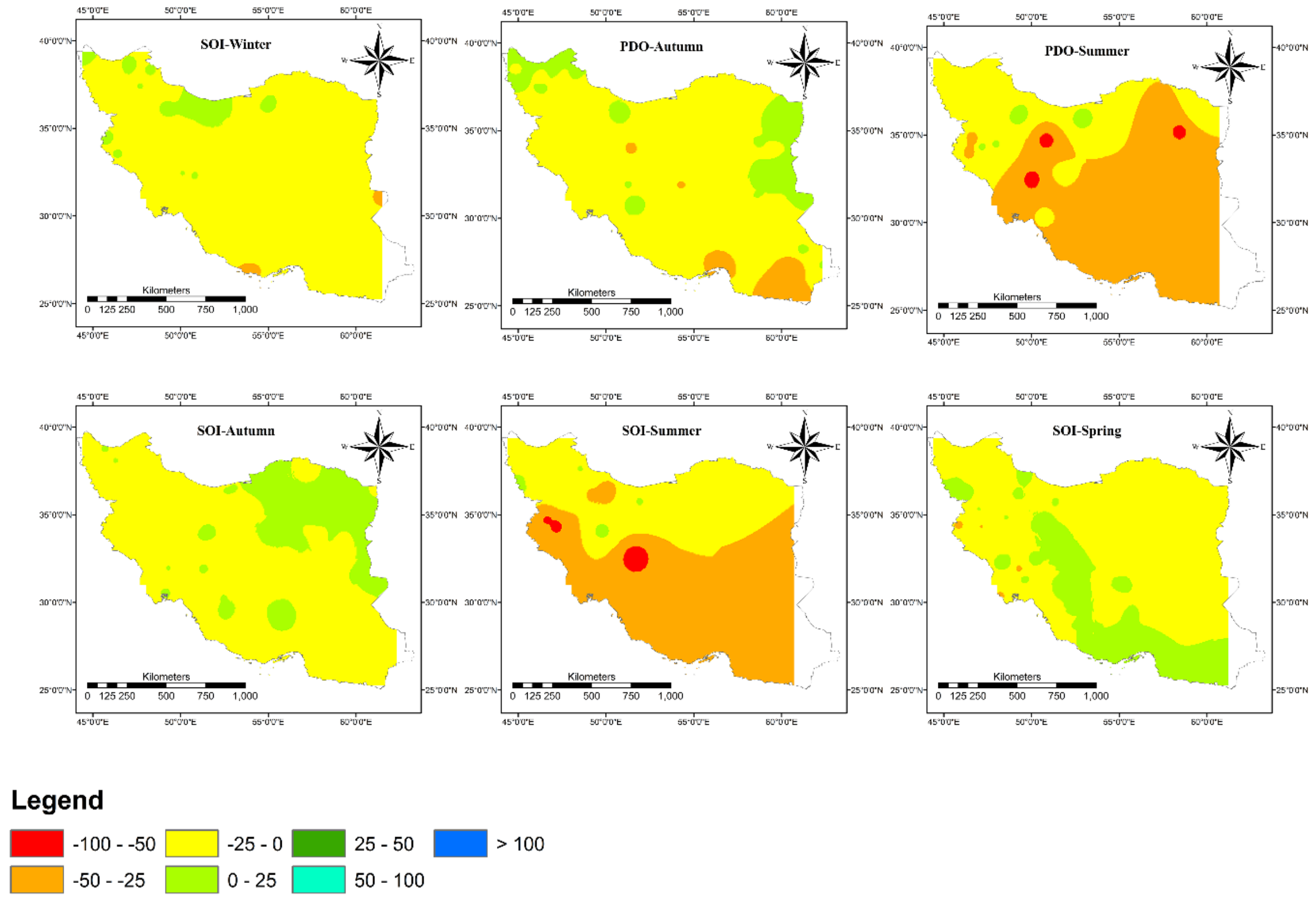

**Figure 8.** Total change in precipitation due to the effect of SOI, NAO, and PDO in the range of -1 <*x*< 1 where*x* denotes the SOI, NAO and PDO value.

Based on Figure 8. Just NAO causes an increase in precipitation in a small area in the west. In the autumn, much of the country area receives less than average precipitation. NAO, PDO and SOI Just intensify the precipitation in the northwest and west, northeast and northwest and northeast, respectively.

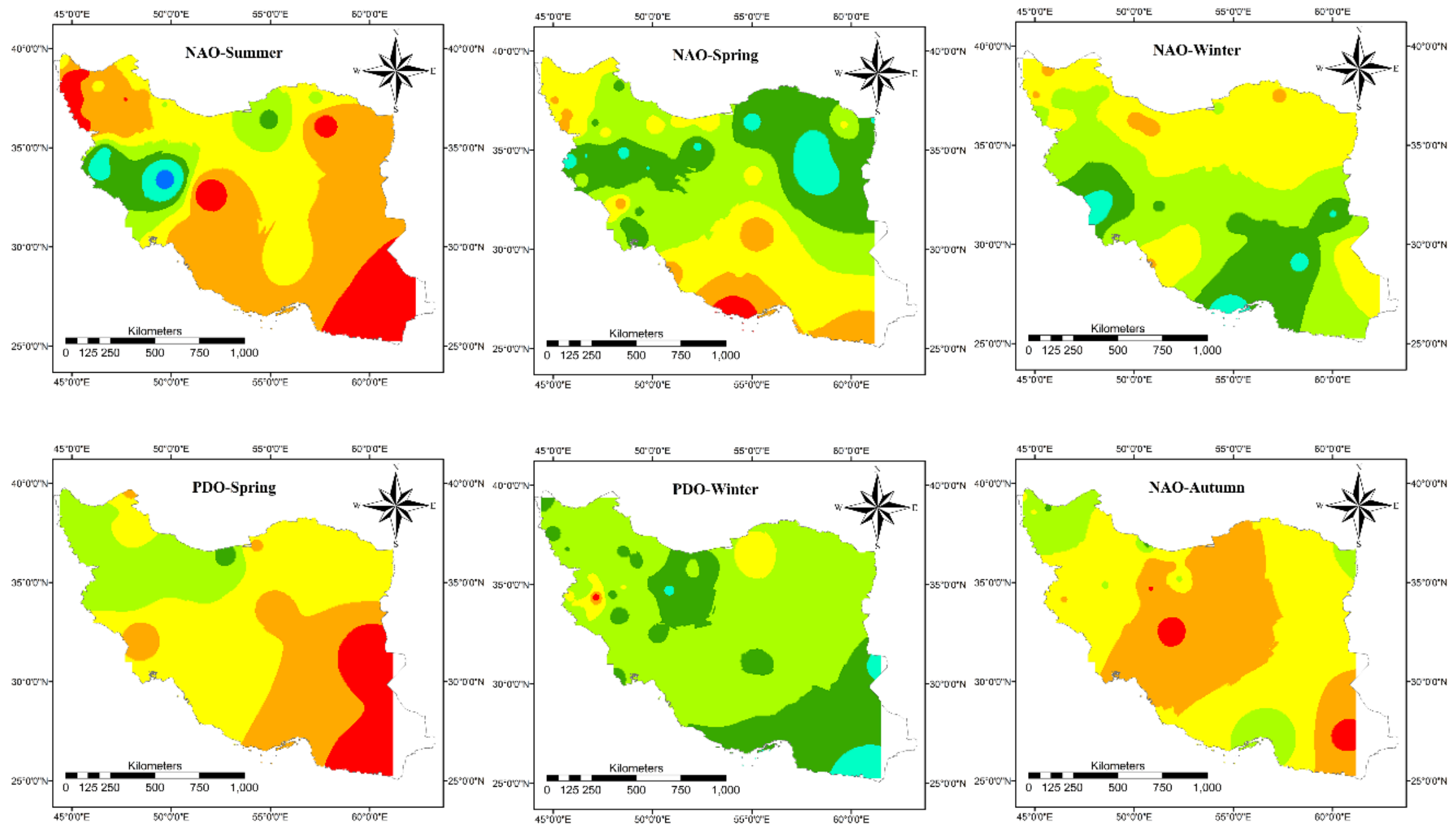

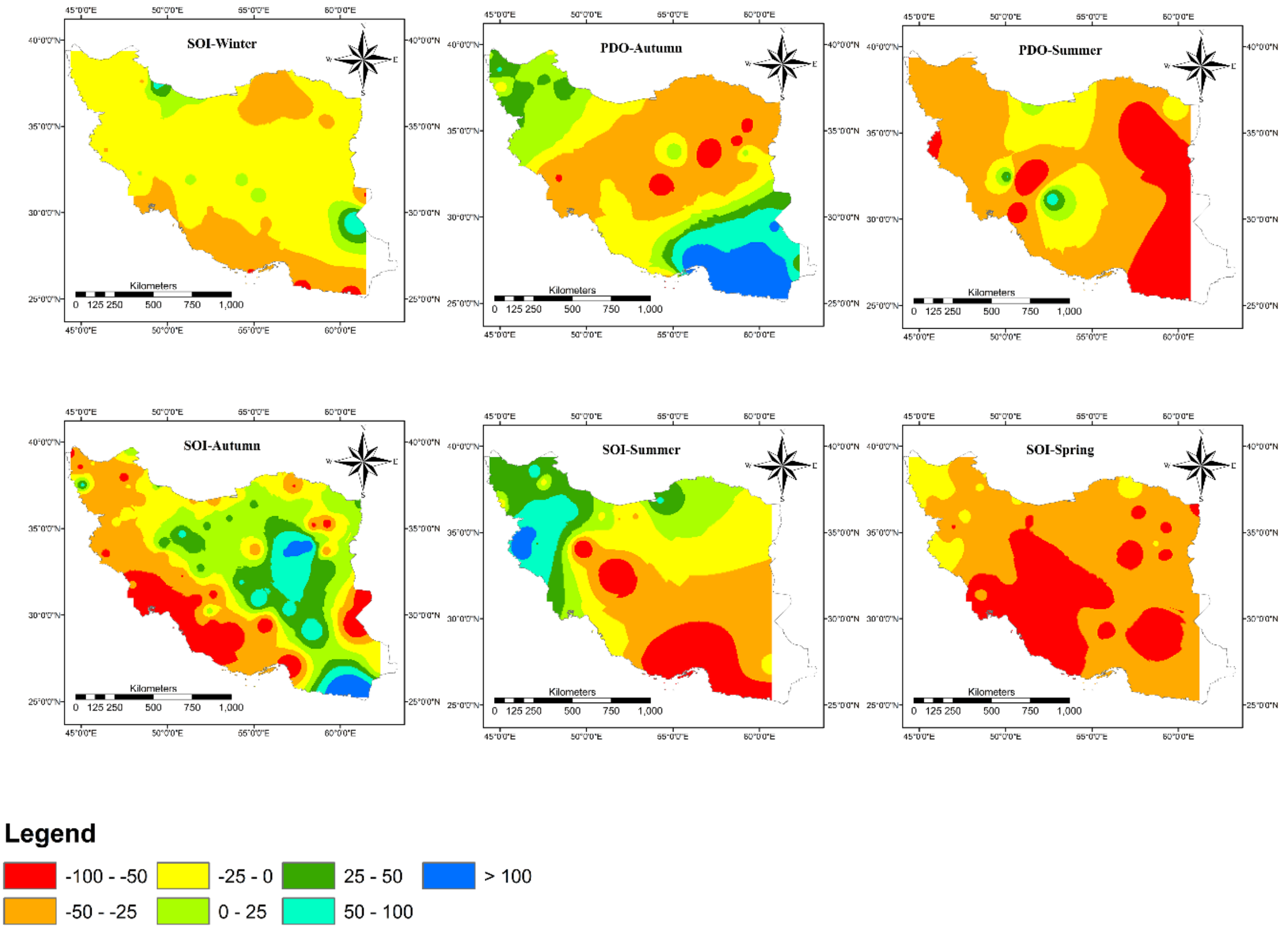

**Figure 9.** Total change in precipitation due to the effect of SOI, NAO, and PDO in the range of 1 <*x*< 2 where*x* denotes the SOI, NAO, and PDO value.

Figure 9 shows the changes in precipitation when the NAO, PDO, and SOI are in the range of 1 <*x*< 2. In winter and spring, NAO in this range intensifies the precipitation in the southern and northern half of the country, respectively while the severity of this intensification in the spring is higher. The rest of the country experiences less than the average precipitation. In the summer and autumn, reducing the precipitation is dominant in the whole country except in the summer in a small area in the west. PDO increases the precipitation in the entire of the country in the winter and northwest and especially southeast in the autumn.

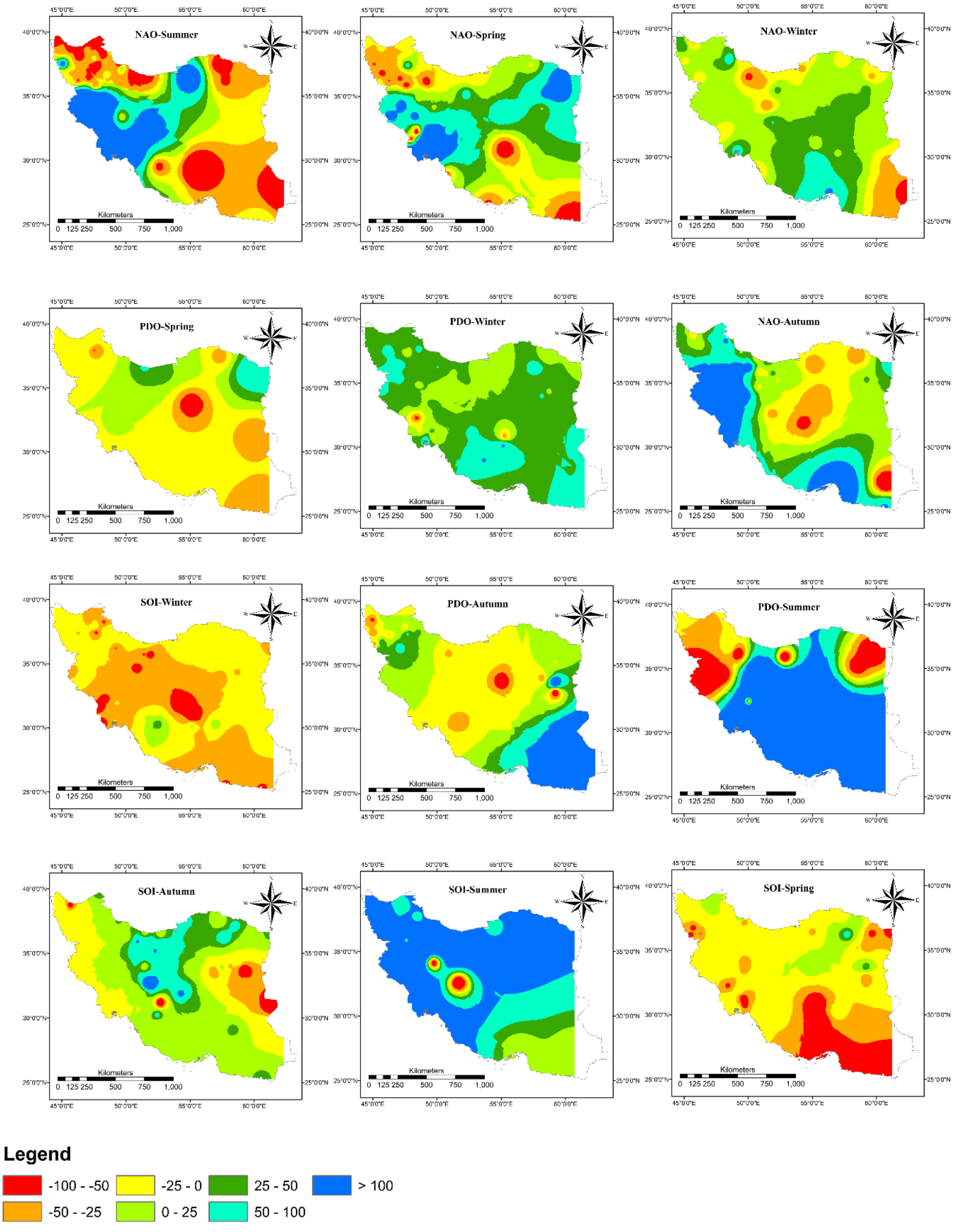

**Figure 10.** Total change in precipitation due to the effect of SOI, NAO, and PDO in the range of $x<2$ where $x$denotes the SOI, NAO and PDO value.

The middle part of the country experiences a 50% decrease in precipitation in the autumn. This is a dominant pattern in the summer in the whole country and with a lower intensity in the spring except in the northwest areas. Generally, SOI decreases the winter and spring precipitation by about 25% while it reaches to above 50% in the southwest and middle of the country. In the summer, northwest and northwest experience above-average precipitation whiles the middle and southeast of the country suffer from about 50% decrease in precipitation. In the autumn, the precipitation decreases about 50% to 100% in the west and southwest while the other parts of the country experience an increase in precipitation.

Finally, Figure 10 shows the total changes in precipitation when the NAO, PDO, and SOI are in the range of $x > 2$. In the winter, when NAO is in this range, the whole parts of the country except north and southeast areas experience at least 25% increase in precipitation. Also, PDO intensifies the precipitation in almost all parts of the country. In contrast, SOI in this range decreases the precipitation by about 25%, and in some areas, it reaches to 100%. NAO in this range decreases the precipitation in the northwest and southeast while the other parts of the country the precipitation increase, especially in the west and northeast areas. The PDO reduces precipitation in the spring in the country except in the north and northeast. Also, the SOI reduces precipitation in almost all areas in the country, and it reaches the highest amount in the southeast. PDO and SOI have virtually the same increasing impact on precipitation in the country in summer except in the northwest and northeast when the PDO is in this range. However, NAO decreases the precipitation considerably in the east and northwest of the country while the other areas experience increases in precipitation. In the autumn, NAO reduces the precipitation in the middle of the country and some areas in the southeast. Other parts of the country receive more than the average precipitation. PDO intensifies the precipitation in the east and northwest areas, and the other areas suffer from decreasing the precipitation. Finally, SOI increases the precipitation in the country except in western border areas and some areas in the east. Based on figures 4 to 8, it can be concluded that SOI, PDO, and NAO have a substantial impact on seasonal precipitation in Iran. This finding is in general agreement with previous studies, e.g., [21,22,25]. However, these studies mostly focused on autumn and winter precipitation and just SOI index was used in their studies.

## 5. Conclusions

In this research, the impact of large-scale circulation drivers, including SOI, PDO, and NAO on seasonal precipitation was investigated. For this purpose, 103 synoptic stations with at least 30 years of data were considered. The highest correlation between these indices in the previous 12 months and seasonal precipitation was calculated. The month with the highest impact on seasonal precipitation for each index was determined. Finally, the amount of decrease and increase in precipitation due to these indices in five classes was calculated. Based on the results, NAO and SOI have a positive correlation with precipitation in the autumn, winter, and spring in most stations while they have a negative correlation with summer precipitation. This pattern is the same for PDO in winter and spring while there is a positive and negative correlation with the summer and autumn precipitation, respectively. Also, almost all correlations lie in the range of 0.6 to -0.6, and in rare cases, this correlation reaches a value out of this range. SOI, NAO, and PDO have the most substantial impact on seasonal precipitation, respectively. In the autumn, winter, and spring which most of the precipitation occurs, SOI has a significant correlation with more than 70% of the stations in these three seasons. Also, NAO has a significant correlation with about 40% and PDO, with 30% of the stations in general. The most meaningful correlation for SOI is with autumn precipitation and for NAO and PDO with winter precipitation. The month in which these indices have the highest correlation with seasonal precipitation does not follow a distinctive pattern. However, it is possible to conclude that the areas around the Zagros and Alborz Mountain chains almost have the highest correlation with an especial month.

SOI has an adverse impact on changes in precipitation compared to the PDO and NAO in winter. El Niño and strong El Niño intensify the winter precipitation while PDO in the warm and very warm phase and NAO in the cold and very cold phase decrease the precipitation in almost all areas. In contrast, the La Niña and strong La Niña decrease the winter precipitation while PDO in the cold and very cold phase and NAO in the warm and very warm phase increase the precipitation in almost all areas. In the other seasons, they do not follow a unique pattern, and each one has a special impact on precipitation in different parts of the country. In the neutral phase, all these indices have almost the same impact on seasonal precipitation, and in most cases, it is decreasing. In all seasons, there is no area with more than 100% decrease in precipitation, however; there are several areas with more than 100% increase in precipitation due to these indices impact.

While the results were promising, there are some limitations. The accuracy of correlation analyses depends on the sample size. The longer data period leads to the better correlation analyses. In the next step, it is needed to characterize the ocean-atmospheric interaction with precipitation. Also, it is worthy to note that a correlation does not automatically mean that one variable causes the change in the values of another variable. Correlation analysis just examines the statistical relationship and does not show the physical mechanism between the variables.

The results of this research can be used effectively for monitoring and managing the water resources in Iran, which is located in an arid and semi-arid region with limited water resources. Knowing the increase and decrease in precipitation in different areas can be used to manage the drought and flood in different areas. As an example, the correlation between the winter and spring precipitation with these indices in southwest areas shows that the strong El Niño intensifies the precipitation considerably. This happens in the winter and spring of 2019 in this area which leads to considerable floods in this region. In this research, the overall impact of these indices on seasonal precipitation was investigated. It means that the authors considered the general impact of these indices on the seasonal precipitation, and it is possible to find some cases with a different impact compared to the results of this research. To clarify this, it is needed to perform a study that investigates the probability of an increase or decrease in precipitation associated with these indices. Also, in that case, it is possible to calculate the amount of increase or decrease in precipitation.

**Author Contributions:** Conceptualization, Majid Dehghani and Somayeh Salehi; methodology, Majid Dehghani; software, Majid Dehghani; validation, Somayeh Salehi, Amir Mosavi. and Shahaboddin Shamshirband; formal analysis, Majid Dehghani and Somayeh Salehi; investigation, Majid Dehghani and Somayeh Salehi; resources, Shahaboddin Shamshirband, Pedram Ghamisi; data curation, Majid Dehghani and Somayeh Salehi; writing—original draft preparation, Majid Dehghani; writing—review and editing, Amir Mosavi, Pedram Ghamisi; visualization, Majid Dehghani and Somayeh Salehi; supervision, Shahaboddin Shamshirband, Pedram Ghamisi.

**Conflicts of Interest:** The authors declare no conflict of interest.

## Appendix A

**Table A1**. Annual average of precipitation in synoptic stations during 1987–2016

| Station Name | Annual average (mm) | X (degree) | Y (degree) | Elevation (m) | Station Name | Annual average (mm) | X (degree) | Y (degree) | Elevation (m) |
|---|---|---|---|---|---|---|---|---|---|
| Abadan | 150 | 48.25 | 30.37 | 7 | Ghoochan | 314 | 58.50 | 37.07 | 1287 |
| Abade | 129 | 52.67 | 31.18 | 2030 | Gonabad | 129 | 58.68 | 34.35 | 1056 |
| Abali | 546 | 51.88 | 35.75 | 2465 | Gorgan | 515 | 54.27 | 36.85 | 13 |
| Abumoosa | 127 | 54.83 | 25.83 | 7 | Hamedan | 301 | 48.53 | 34.87 | 1742 |
| Ahar | 285 | 47.07 | 38.43 | 1391 | Ilam | 562 | 46.43 | 33.63 | 1337 |
| Ahwaz | 210 | 48.67 | 31.33 | 23 | Iranshahr | 105 | 60.70 | 27.20 | 591 |
| Aligoodarz | 388 | 49.70 | 33.40 | 2022 | Isfahan | 129 | 51.67 | 32.62 | 1550 |
| Anar | 69 | 55.25 | 30.88 | 1409 | Islam abad | 462 | 46.47 | 34.12 | 1349 |
| Arak | 298 | 49.77 | 34.10 | 1708 | Jask | 115 | 57.77 | 25.63 | 5 |
| Ardabil | 278 | 48.28 | 38.25 | 1332 | Jolfa | 212 | 45.67 | 38.75 | 736 |
| Astara | 1359 | 48.87 | 38.42 | 18 | Kangavar | 384 | 47.98 | 34.50 | 1468 |
| Babolsar | 912 | 52.65 | 36.72 | -21 | Karaj | 246 | 50.90 | 35.92 | 1313 |
| Bam | 56 | 58.35 | 29.10 | 1067 | Kashan | 127 | 51.45 | 33.98 | 982 |
| Bandar Abbas | 171 | 56.37 | 27.22 | 10 | Kashmar | 187 | 58.47 | 35.20 | 1110 |
| Bandar-e- Anzali | 1730 | 49.47 | 37.47 | 26 | Kerman | 128 | 56.97 | 30.25 | 1754 |
| Bandar-e-Lenge | 130 | 54.83 | 26.53 | 23 | Kermanshah | 411 | 47.15 | 34.35 | 1319 |
| Bandar-e-Mahshahr | 197 | 49.15 | 30.55 | 6 | Khalkhal | 370 | 48.52 | 37.63 | 1796 |
| Bijar | 334 | 47.62 | 35.88 | 1883 | Khash | 147 | 61.20 | 28.22 | 1394 |
| Birjand | 151 | 59.20 | 32.87 | 1491 | Khoorbiabanak | 76 | 55.08 | 33.78 | 845 |
| Bojnurd | 252 | 57.32 | 37.47 | 1091 | Khoramabad | 469 | 48.28 | 33.43 | 1148 |
| Boshruye | 86 | 57.45 | 33.90 | 885 | Khoramdareh | 295 | 49.18 | 36.18 | 1575 |
| Bostan | 195 | 48.00 | 31.72 | 8 | Khoy | 262 | 44.97 | 38.55 | 1103 |
| Brojen | 250 | 51.30 | 31.95 | 2197 | Kish | 155 | 53.98 | 26.50 | 30 |
| Bushehr | 255 | 50.83 | 28.98 | 20 | Koohrang | 1336 | 50.12 | 32.43 | 2285 |
| Bushehr (Coastal) | 252 | 50.82 | 28.90 | 8 | Mahabad | 393 | 45.72 | 36.77 | 1385 |
| Chabahar | 117 | 60.62 | 25.28 | 8 | Makoo | 302 | 44.43 | 39.33 | 1411 |
| Dezful | 313 | 48.42 | 32.27 | 83 | Maragheh | 279 | 46.27 | 37.40 | 1478 |
| Dogonbadan | 420 | 50.77 | 30.43 | 700 | Mashhad | 242 | 59.63 | 36.27 | 999 |
| Fassa | 285 | 53.68 | 28.97 | 1288 | Masjed Soleyman | 399 | 49.28 | 31.93 | 321 |
| Ferdos | 128 | 58.17 | 34.02 | 1293 | Minab | 206 | 57.08 | 27.10 | 30 |
| Garmsar | 111 | 52.27 | 35.20 | 825 | Miyaneh | 276 | 47.70 | 37.45 | 1110 |
| Ghaen | 161 | 59.17 | 33.72 | 1432 | Nehbandan | 121 | 60.03 | 31.53 | 1211 |
| Gharakhil | 718 | 52.77 | 36.45 | 15 | Noushahr | 1303 | 51.50 | 36.65 | -21 |
| Ghazvin | 312 | 50.05 | 36.25 | 1279 | Omidiye | 256 | 49.67 | 30.77 | 27 |
| Ghom | 137 | 50.85 | 34.70 | 877 | Pars Abad | 276 | 47.92 | 39.65 | 32 |

**Table 1A.** Continued

| Station Name | Annual average (mm) | X (degree) | Y (degree) | Elevation (m) |
|---|---|---|---|---|
| Piranshahr | 646 | 45.13 | 36.67 | 1455 |
| Ramhormoz | 292 | 49.60 | 31.27 | 151 |
| Ramsar | 1227 | 50.67 | 36.90 | -20 |
| Rasht | 1322 | 49.65 | 37.20 | 37 |
| Ravansar | 500 | 46.65 | 34.72 | 1380 |
| Sabzevar | 185 | 57.72 | 36.20 | 978 |
| Sad-e- Drudzan | 458 | 52.43 | 30.22 | 1620 |
| Saghez | 445 | 46.27 | 36.25 | 1523 |
| Sanandaj | 378 | 47.00 | 35.33 | 1373 |
| Sar Pol Zahab | 411 | 45.87 | 34.45 | 545 |
| Sarab | 239 | 47.53 | 37.93 | 1682 |
| Sarakhs | 191 | 61.17 | 36.53 | 235 |
| Saravan | 106 | 62.33 | 27.33 | 1195 |
| Sardasht | 823 | 45.50 | 36.15 | 1670 |
| Semnan | 135 | 53.55 | 35.58 | 1131 |
| Shahr-e-Babak | 151 | 55.13 | 30.10 | 1834 |
| Shahrkord | 320 | 50.85 | 32.28 | 2049 |
| Shahrud | 152 | 54.95 | 36.42 | 1345 |
| Shamiran | 414 | 51.62 | 35.78 | 1976 |
| Shargh-e-Isfahan | 103 | 51.87 | 32.67 | 1543 |
| Shiraz | 323 | 52.60 | 29.53 | 1484 |
| Sirjan | 139 | 55.68 | 29.47 | 1739 |
| Tabas | 78 | 56.92 | 33.60 | 711 |
| Tabriz | 246 | 46.28 | 38.08 | 1361 |
| Takab | 318 | 47.12 | 36.38 | 1765 |
| Tehran | 228 | 51.38 | 35.73 | 1419 |
| Torbat-e-Heydariyeh | 253 | 59.22 | 35.27 | 1451 |
| Urmia | 302 | 45.08 | 37.53 | 1316 |
| Yasuj | 797 | 51.68 | 30.83 | 1832 |
| Yazd | 51 | 54.28 | 31.90 | 1237 |
| Zabol | 54 | 61.48 | 31.03 | 489 |
| Zahedan | 76 | 60.88 | 29.47 | 1370 |
| Zanjan | 284 | 48.48 | 36.68 | 1663 |